%% file: draft.tex
\documentclass[sigconf]{acmart}

\usepackage{booktabs} 
\usepackage{graphicx}
\usepackage{subcaption}

\begin{document}
\title{Deep neural network marketplace recommenders in online experiments}

\author{Simen Eide}
\affiliation{%
  \institution{Schibsted Media Group}
  \city{Oslo}
  \country{Norway}
}
\email{simen.eide@schibsted.com}

\author{Ning Zhou}
\affiliation{%
  \institution{Schibsted Media Group}
  \city{Oslo}
  \country{Norway}
}
\email{ning.zhou@schibsted.com}

\renewcommand{\shortauthors}{S. Eide and N. Zhou}

\begin{abstract}
Recommendations are broadly used in marketplaces to match users with items relevant to their interests and needs. To understand user intent and tailor recommendations to their needs, we use deep learning to explore various heterogeneous data available in marketplaces. This paper focuses on the challenge of measuring recommender performance and summarizes the online experiment results with several promising types of deep neural network recommenders - hybrid item representation models combining features from user engagement and content, sequence-based models, and multi-armed bandit models that optimize user engagement by re-ranking proposals from multiple submodels. The recommenders are currently running in production at the leading Norwegian marketplace \textit{FINN.no} and serves over one million visitors everyday.

\end{abstract}

%
%



\copyrightyear{2018} 
\acmYear{2018} 
\setcopyright{acmlicensed}
\acmConference[RecSys '18]{Twelfth ACM Conference on Recommender Systems}{October 2--7, 2018}{Vancouver, BC, Canada}
\acmBooktitle{Twelfth ACM Conference on Recommender Systems (RecSys '18), October 2--7, 2018, Vancouver, BC, Canada}
\acmPrice{15.00}
\acmDOI{10.1145/3240323.3240387}
\acmISBN{978-1-4503-5901-6/18/10}

\maketitle

\addtolength\belowcaptionskip{-1mm}

\input{body-conf}

\bibliographystyle{ACM-Reference-Format}
\bibliography{bibliography}

\end{document}

%% file: body-conf.tex
\section{Introduction}

\label{sec:intro}
Marketplaces are platforms where users buy and sell various types of items. The items can range from low-value ones such as books and clothes to high-value ones such as cars and real estate properties. Sellers can also post non-tangible items such as job openings and services. Many marketplace sellers are non-professional individuals selling used items, therefore marketplaces can be viewed as a special type of e-commerce that involves unique items across multiple categories from a very large and fragmented seller group. 

\begin{figure}
\centering
\begin{subfigure}{0.235\textwidth}
	\includegraphics[width=\linewidth]{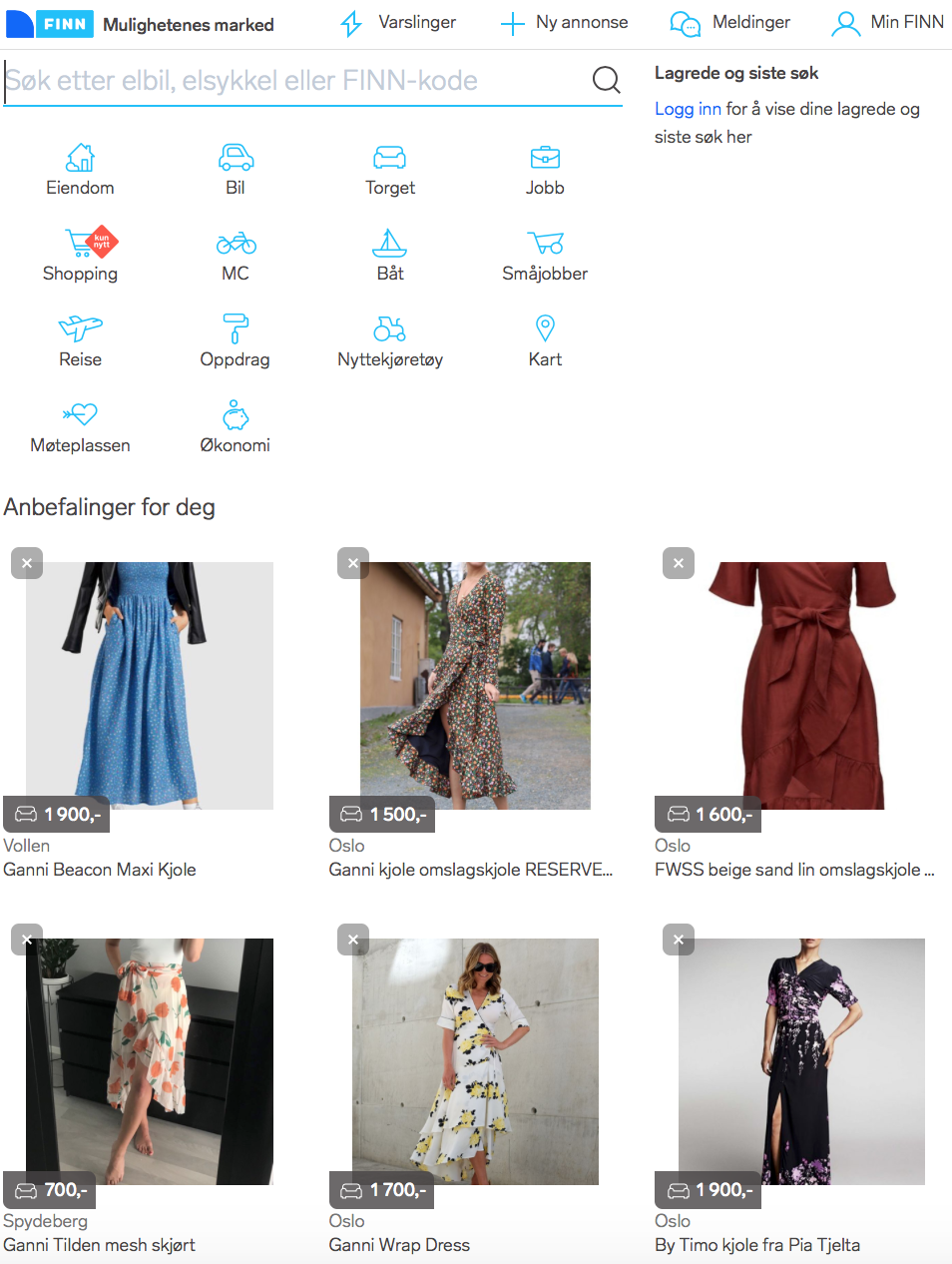}
    \caption{Item feed}
\end{subfigure}
\begin{subfigure}{0.215\textwidth}
	\includegraphics[width=\linewidth]{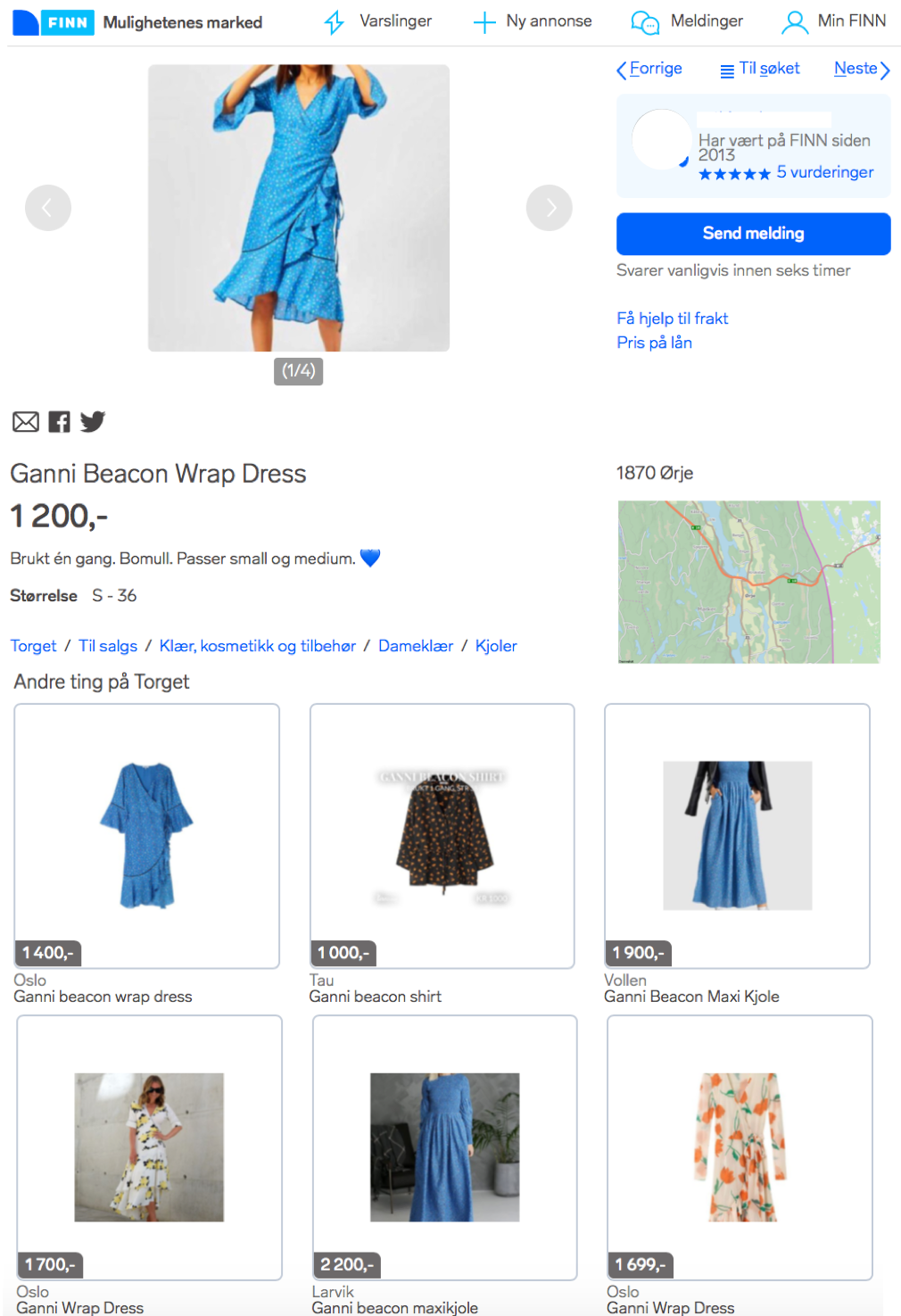}
    \caption{Similar item}
\end{subfigure}
\caption{Sample marketplace recommendations.}
\label{fig:marketplace}
\end{figure}
 
Recommendation systems are broadly used in marketplaces to match buyers with items relevant to their interests and needs. It is more challenging than the standard e-commerce product recommendation due to the following reasons: 
(1) Marketplace items are often secondhand and therefore in a sense unique. This results in a short lifespan for a specific item. Thus cold-start issues are very prominent.
(2) The items are often poorly described by unstructured data, making conventional cold-start algorithms harder to apply.
(3) The interaction between buyers and sellers is often tricky to track, as it can happen outside the platform. Transactions cannot always be confirmed, so the recommender system must be able to change focus from sold items quickly to remain relevant.

There is already substantial research on recommender systems validated on offline metrics. In this paper, we describe three new marketplace recommenders and benchmark them through online experiments against industry standard models such as matrix factorization to test their performance in a production environment where short lifespan and low quality content are common features of the items recommended.
The three models we describe here are:
(1) A hybrid item-item recommender that utilizes behavior data, image and text to build a robust item representation against cold-start issues.
(2) A sequence based user-item recommender that is time-aware and can utilize the hybrid item representation above to quickly build a user's profile.
(3) A higher-level multi-armed bandit algorithm that prioritizes between multiple submodel recommendations into a personalized item feed. It allows the feed to cover both long and short term user interests.

The production environment for testing is the leading Norwegian marketplace \textit{FINN.no}. It has over one million active items in 10+ categories and 200+ subcategories for sell, and serves over one million visitors per day. There are two recommendation features at \textit{FINN.no}: an item feed on the frontpage and a similar item recommendation widget on the item detail page. One sample of each is shown in Figure \ref{fig:marketplace}. The hybrid model is used for the similar item recommendation, while the rest two are used for the item feed.

\begin{figure*}
\centering
\begin{subfigure}{0.6\textwidth}
	\includegraphics[width=\linewidth]{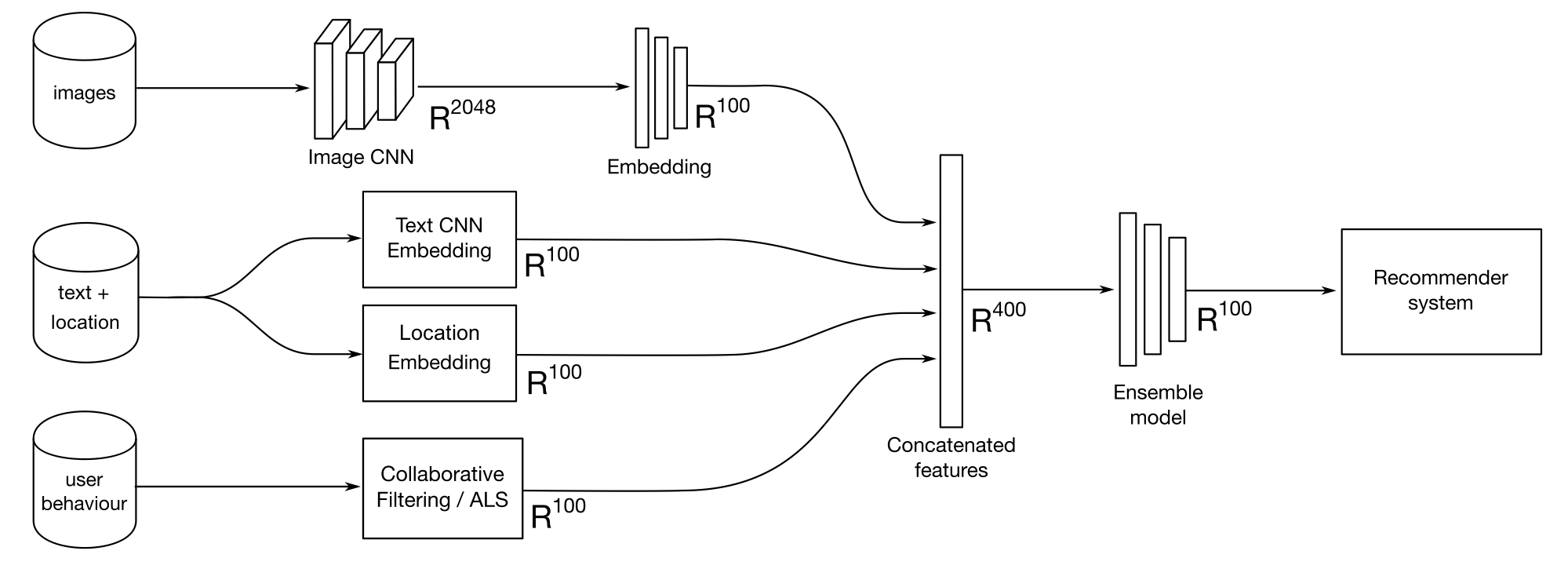}
	\caption{Hybrid item representation model.}
    \label{fig:hybrid}
\end{subfigure}
\begin{subfigure}{0.35\textwidth}
	\includegraphics[width=\linewidth]{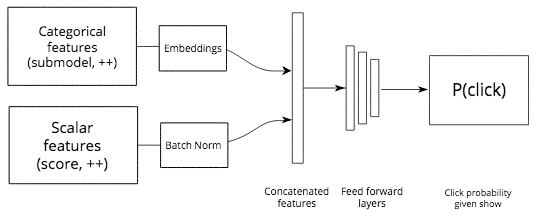}
	\caption{Deep multi-armed bandit model.}
    \label{fig:bandit}
\end{subfigure}
\caption{Overview of the recommender models.}
\label{fig:overview}
\end{figure*}

\section{Related works}
\label{sec:related}

Search and discovery are the two main ways for users to find relevant items on a content platform. Using recommendations to improve the discovery experience has been a hot topic in recent years. Both collaborative filtering \cite{barkan2016item2vec} \cite{sarwar2001cf} and content based methods \cite{Christidis2012ATR} \cite{grbovic2015commerce} are commonly used in item similarity ranking for e-commerce. Recent works like LightFM \cite{Kula15:lightfm} combine the two to address the cold-start problem in recommendations. The state-of-the-art recommenders such as \cite{Criteo:2016} and \cite{pinterest:2017} often use learning-to-rank to model from a complex set of features including text, image, user profile, behavior, etc. This strategy is effective with marketplaces recommenders as well \cite{Eide18}. Moreover, models of cascade \cite{liu2017cascade} or sequential attention \cite{atrank:2018} that consider a longer user behavior history show particularly good results. Our sequential recommender draws inspiration from \cite{deboom17} and shows positive performance improvements in online experiments. Multi-armed bandits for prioritizing among multiple sources is applied in different use cases. For example, Pinterest \cite{StephaniedeWet17} uses a blending algorithm to generate a mixed feed of their recommendation content and social content. We compared several multi-armed bandit models to leverage the behavior-content complementation in the marketplace scenario.

Recommenders are usually evaluated through click and conversion rates from online controlled experiments \cite{Kohavi13}. There are offline metrics such as predicted ratings and ranking correlation \cite{Gunawardana09} that can be used with lower preparation cost, and serve as a proxy to the online tests.

\section{Experimentation platform}
\label{sec:platform}

A robust experimentation platform for production requires both a fast iteration cycle from idea to production and a risk-averse test procedure to secure the product quality. From the business perspective, the goal of a recommendation system is to increase the amount of conversions, so we choose to rely on A/B tests to select good candidate models based on conversion metrics in production.

Our experimentation system consists of offline and online experiments. We use the offline ones mainly to quickly rule out the candidate models that are clearly underperforming and therefore not feasible for online testing. This allows us not only to utilize online experimentation resources more efficiently, but also to lower the risk of poor user experience. We use the offline metric Hit Rate@n ($HR@n$) \cite{Deshpande04} as a proxy for online conversion metrics. Historical data are splitted into a training set $D_{t<t_{th}}$ and a test set $D_{t{\geq}t_{th}}$ by a specific date $t_{th}$. A model is trained on $D_{t<t_{th}}^{u}$ and recommends top $n$ items for each user $u$. $HR@n$ is defined as the average per user hits of the top $n$ items found in $D_{t{\geq}t_{th}}^{u}$. 

The candidates that perform better or in the same range as the baseline model in offline tests are deployed online to A/B test in production with gradually higher traffic. We generally follow \cite{Kohavi14} for the experiment procedure design. Online experiments are evaluated with click through rate and other conversion metrics such as seller contacting rate. We monitor all experiments for significance using a binomial test. During an A/B test, we also consider the stability of the result over time. In addition to aggregating all the samples to calculate the final test result, we monitor the model performance in smaller time bins during the whole test. If model B is not better than model A, the probability of model B performing better than model A in almost all time bins will be very small. When model B shows improvement but not significant, we gradually increase its traffic up to 50\% in a longer A/B test to see if it will reach significance. 

\section{Hybrid item representation model}
\label{sec:hybrid}

Similar to Lightfm\cite{Kula15:lightfm}, we combine content-based features with user behaviors to solve the cold start challenge of collaborative filtering. We design a hybrid model mixing item representations from user behavior, text, image and location to find similar items. As shown in Figure \ref{fig:hybrid}, we first train the item representations independently and then ensemble them using a separated model. 

The individual features are generated in the following way:
(i) The behavior-based item representation is from matrix factorization implemented with the industry standard Alternating Least Squares (ALS) \cite{spark:als}. The training data consists of 20-day look-back of item clicks and conversion signals such as contacting seller. The conversion signals are stronger and therefore acquired more weights. All hyper-parameters are tuned through offline and online experiments.
(ii) The textual features are from the unstructured free text in the item title and description. We first train an item category classifier to map the text to the seller-assigned category, and then extract the top layer of this classifier as the textual item representation. The classifier is a variant of the Convolutional Neural Network (CNN) architecture of \cite{collobert2011natural} using a word2vec \cite{mikolov2013word2vec} model pre-trained on marketplace corpus.
(iii) The image features are generated by training a model to predict the item title from its image. The title is projected into word embeddings by the word2vec model mentioned above. We use the title instead of the category to classify the images. This gives us a richer feature space. For example, it enables the item representation to distinguish between "wedding dress" and "summer dress", despite both belong to the same category "dresses". The model uses the penultimate layer of a pre-trained \textit{Inception-v3} model \cite{szegedy2016rethinking} and stacks seven linear feed-forward layers on top to predict the title in the word embedding space. It is then trained by minimizing the mean squared error between the predicted title embeddings and the real title embeddings.
(iv) We do not use the simple geographical distance to represent location, because it can be misleading. There are hidden factors such as population density and ease of transport that affect the impact of location. We train the location representation based on the historical user behavior, since the items a user showed interest in implicitly tell us a lot about the hidden factors. Similar to (i), we factorize a user-postcode matrix and use the postcode embedding as the location representation.

The different item representations are merged into a hybrid one using a Siamese network. All the item representations are concatenated and passed through an attention layer. The attention layer allows the model to focus more on textual and image features if the collaborative filtering features are missing, and vice versa. Then, a towering feed-forward network compresses the features into a 100-dimensional item representation due to the capacity limit of our serving infrastructure. Two items are compared using the cosine similarity of this representation. The training data consists of co-converted item pairs, i.e. two items that get conversions from the same user on the same day, and negative sampling item pairs that are unlikely to co-convert. The underlying assumption is that co-converted items are likely to be similar. 

\section{Sequence-based models}
\label{sec:seq}

Sequence-based models look into a user's click history and predict what items they will click next. In contrast to matrix factorization, such models are time-aware and can take recent clicks into account more in the predictions. They take the $n$ most recent items clicked by a user and project them into the item space defined by the hybrid item representation described in Section \ref{sec:hybrid}. The item sequence is fed through a recurrent neural network. At every step $t$, the models use the click history sequence $\{x_{t-n}, ..., x_{t}\}$ to predict a sequence $\{\hat{x}_{t+1}, ..., \hat{x}_{t+k}\}$ of the future $k$ steps. The accuracy is calculated by the cosine similarity of the predicted sequence $\{\hat{x}_{t+1}, ..., \hat{x}_{t+k}\}$ at $t$ and the actual clicked sequence $\{x_{t+1}, ..., x_{t+k}\}$ at $t+k$.

In \cite{deboom17}, two-layer GRU was shown to be the optimal architecture. We also tested different variants using LSTM or GRU, adding additional stacked recurrent layers, or adding attention to the outputs. However, we did not observe any additional significant improvements over using one straight forward GRU layer.

\section{Multi-armed bandit models}
\label{sec:feed}
 
The multi-armed bandit models can generate a live item feed for a specific impression. The bandits are not recommenders by themselves but re-rankers that receive proposals from independent submodels as input and re-rank all the proposed items from most to least relevant by estimating their click probabilities with a value function. A submodel can be a recommender of any type, e.g. sequence-based or matrix factorization, that returns its top proposals along with their corresponding scores. Typically, a bandit is connected to 6-10 submodels. In order to avoid local minimal during training, we adopt a simple epsilon-greedy policy and add 5\% random items in every recommended list to the bandit.

This approach differs from the ones such as \cite{Covington2016} that use simple models as the first filter into a more complex model. Instead, the bandit enables us to directly leverage the results from well-performing models tuned for different scenarios. It utilizes the scores from each submodel and focuses on evaluating submodel reliability as well as contextual information such as visiting time, device, landing page type (main frontpage or categorical frontpage), etc.

The row-separated feed described in Section \ref{sec:rowfeed} is just a naive baseline for experimentation purpose. We propose two different value functions to estimate click probabilities: \textit{regression bandit} in Section \ref{sec:regbandit} and \textit{deep classification bandit} in Section \ref{sec:deepbandit}. The former is selected mainly due to its good interpretability, whereas the latter allows us to increase value function complexity and get a better online performance. 

\subsection{Row-separated feed}
\label{sec:rowfeed}
This simple baseline has a fixed number of rows and serves one submodel per row. The row order is also fixed based on the individual performance of each submodel. We choose this setup because it is commonly used by many e-commerce recommendation plug-ins.

\subsection{Regression bandit}
\label{sec:regbandit}
The regression bandit, as its name, using regression to approximates the click probability based on submodel score, submodel type and contextual information. Submodel scores are binned into 10 buckets. The remaining features are categorical and one-hot encoded. The recommendation impressions are grouped across these features and the target value is the average number of clicks per group. A ridge regression model is fit on the encoded features, as the weights of the unregularized versions often explode.

\subsection{Deep classification bandit}
\label{sec:deepbandit}
The deep classification bandit estimates the click probability as a classification problem. This allows us to use more complex functions taking in a larger set of features than the regression bandit.

The model input is a mix of scalars (submodel score, item position, hour of day, etc.) and categorical variables (submodel type, location, device, weekday, etc.). The scalars are normalized through batch normalization, and the categorical variables are one-hot encoded. A towering feed-forward network is applied to the features and finally outputs a click probability. The dataset contains an unequal ratio of click v.s. view events (i.e. only a small ratio of recommended items were clicked) that can cause severe instabilities in training. This is solved by introducing class weights in the loss function and applying L2 regularization on the model weights.

\section{Results and discussions}
\label{sec:observations}

We conducted online experiments on both the similar item widget and the item feed shown in Figure \ref{fig:marketplace}. Similar item widget is shown at the bottom or the sidebar of an item detail page, assuming that users view the item details because they are interested in this item and therefore interested in similar items. The item feed is usually displayed at the main and categorical frontpages of the marketplace, aiming to provide new and relevant items for users. We choose to test the hybrid item representation in the similar item widget while the multi-armed bandits and sequence-based model in the item feed, because the item-item type of models are not capable to create the serendipity experience desired in the item feed scenario \cite{Kaminskas16}.

The experiment results are summarized in Table \ref{tab:results}. The absolute click thorough rate (CTR) varies a lot due to the seasonal effect, new feature launches, etc. Hence we report the more stable CTR improvement $\Delta CTR = (CTR_B - CTR_A) / CTR_A$ when comparing experiments from different periods. By default, the A/B tests last for one week to avoid seasonal impact of weekends and accumulate around one million impressions of the similar item widget and 5-10 million impressions of the item feed. For simplicity, we only show the results of the best of each type after hyper-parameter tuning.

In the similar item scenario, the pure content-based item representation model is 7.7\% worse than the matrix factorization baseline, but the hybrid model outperformed the baseline by 23.8\%. This indicates that the user behavior signals give more relevant recommendations than content-based features but still suffer a lot from cold start. Throughout the experiments, we observed many examples where the hybrid model used content-based features on items with very few clicks, whereas it gave similar results to the matrix factorization approach for items with abundant clicks.

Sequence-based models generally outperformed matrix factorization in the online experiments. When $n=1$, they collapse into an item-item recommender making predictions only using the last seen item. We require $n>1$ in order for the models to generalize and find out that a $n=15$ look-back with a $k=5$ prediction horizon gives the best performance of 21.2\% CTR improvement.

The multi-armed bandit models were tested in two groups with different baselines, because we introduced a major user interface (UI) redesign in between. When we tested the regression bandit against the row-separated feed baseline, we observed 50.1\% CTR improvement. We attribute the improvement mainly to that the regression bandit can mix submodels across the rows and not constrained by the fixed order. An interesting observation is that the scores from matrix factorization submodels do not have a monotonous increasing relationship with CTR. In our experiments, CTR keeps increasing before scores reach around $0.8$, but then falls quickly after that. Investigations show that these ultra high score items have some viral tendencies and do not reflect personal taste, therefore do not transform into clicks. We introduce a break point in the regression to allow the click probability estimation to fall after a threshold, and this gained another 5\% CTR improvement. The bandit also allowed us to add more submodels to the feed and opened up the UI flexibility to increase the number of items recommended and submodels connected. After redesigning the UI, CTR increased even further. The two main drawbacks of the (linear) regression bandit are that it cannot estimate the observed non-linear relationship between submodel scores and CTR, and it is not personalized. The regression function cannot handle the variable dimension explosion by introducing per user aggregation. 

We observed around 10\% CTR improvement from testing the deep classification bandit against the regression bandit. The deep model is personalized, for it is capable to use more features including user embeddings.

During the experiments, we also tried several promising approaches but did not succeed:
(1) We tried factorization machines to solve the cold start problem, but the engineering cost was quite high and did not show significant CTR improvement, so we ended up focusing on hybrid models instead. 
(2) We tried to train models with only strong signals such as repeating visits and messaging, but those models turned out to perform worse due to the low data volume in a short look-back time. 
(3) Though both previous works and user studies show that diversity matters a lot for the item feed, when we tried to optimize explicitly with a diversity metric based on category count, we did not get significant improvements. 
(4) Both the regression and the deep classification value functions had large instabilities during training, and a lot of care had to be done to stabilize it for production usage.

\addtolength\belowcaptionskip{-2mm}
\begin{table}
  \caption{Experiment Results.}
  \label{tab:results}
  \centering
  \begin{tabular}{cccc}
    \toprule
    Model Type & Model A & Model B & $\Delta$ CTR \\
    \midrule
    item-item & matrix factorization & content-based &  -7.7\% \\
    \hline
    item-item & matrix factorization & hybrid & 23.8\% \\
    \hline
    user-item & matrix factorization & sequence-based & 21.2\%	 \\
    \hline
    multi-armed & row-separated feed & regression bandit	& 50.1\% \\
    \hline
    multi-armed & regression bandit & deep bandit & 10.0\% \\
    \bottomrule
  \end{tabular}
\end{table}

\section{Conclusions}
\label{sec:conclusions}
In this paper, we propose three new marketplace recommenders - hybrid item representation, sequence-based model, multi-armed bandit, and analyze their performance in online experiments from a production setting. The results demonstrate the effectiveness of combining collaborative filtering and content features for better item representation in cold start and in sequence-based models. We also present a successful use case of bandits in recommendations as a high-level re-ranker on top of other recommenders. These bandits are useful to utilize contextual information and to combine multiple domain-specific recommenders.

For future works, there are still a lot to explore in the above-mentioned models. In general, we have not explored the content features sufficiently. Adding more content features into the item representations should reduce cold-start problems on user or item more effectively. The exploration strategies of the bandit models can be further improved: New submodels still take a long time to catch up. Moreover, clicks and transactions in marketplaces do not happen instantaneously. The transaction of an item often takes place hours or days after an user first views the item, so the "reward" of recommendations can arrive with delay. Defining the models under a reinforcement learning framework may give them an incentive to explore more broadly and provide a better user experience.

\begin{acks}
This project is a team effort from the recommendation team at \textit{FINN.no}. The authors would like to thank Audun M. {\O}ygard, Thorkild Stray, Bj{\o}rn Rustad and Nicola Barbieri for their contribution to this paper. We also thank the anonymous reviewers for their helpful comments.
\end{acks}